\documentclass[reprint,amsmath,amssymb,aps]{revtex4-1}
\usepackage{physics}
\usepackage{graphicx}
\usepackage{dcolumn}
\usepackage{bm}
\usepackage{amssymb}
\usepackage{braket}
\usepackage{mathtools}
\usepackage{comment}
\usepackage{epstopdf}

\begin{document}

\graphicspath{ {./images/} } 
\title{A iterative finite method approach to Kohn-Sham equations of light atoms: Density Functional Theory tutorial for undergraduate students}
\author{Abhishek Joshi$^\dag$, Sinuh\'e Perea-Puente$^\ddag$}
\affiliation{LNMIIT$^\dag$, Jaipur; King's College London$^\ddag$, United Kingdom}
\email{20mph006}

\date{\today}

\begin{abstract}
In this article we are going to study the FEM solution to the Density Functional description of Helium. Solving self-consistently including electron-electron repulsion and exchange-correlation effects. This project will be split in four different consecutive task with different approaches: Numerical Hartree potential (Sect. I) and nuclear potential (Sect. II) for the hydrogen and resolution of Schrödinger equation for hydrogen (Sect. III) and helium (Sect. IV). The code is included in the format \emph{Sinajo} with $n\in{0,1,2,3,4}$ corresponding with supporting material (code) to each task. The format of each section is standard in three cases. Beginning with a theoretical introduction, we mainly explain the code and show the principal results in form of values or graphics. Later on, we try to study the convergence and stability of the method proposed. Exception to this rule can be found in Section II, corresponding to the analytical deduction of a potential term.

\end{abstract}

\maketitle

\section*{Introduction}
From Born-Oppenheimer approximation in which the motion of the nuclei and the electronic cortex is uncoupled, two theoretical approaches are available for atomic simulation: First, the \emph{Hartree-Fock} method consisting in fermionic (antisymmetric) variational procedure in which the N-body problem is simplified using a unique non-local entity. By the other hand, \emph{density functional theory} relies in exchange and correlation potentials, whose analytic solution are not known.\\

In both situations, the quantum fluctuations are omitted and both uses self-consistency as iterative method, offering similar results. In our case, we are going to use the DFT method, in addition with the \emph{Local Density Approximation} (LDA) in which this exchange-correlations terms are only studied in terms of the electronic structure, ignoring nuclear or gradient dependence's.\\
\section{Poisson's equation in radial coordinates}
The first task consists in adapting the program for solving the Poisson's equation already done in the Lecture sessions adapting it to the atomic problem of the hydrogen, in which the Hartree potential is approximated solving the Poisson's like equation (24).\\

Using finite element method (FEM), we will compare the numerical result with the theoretical $V_{Hartree}$ for hydrogen, given by the expression (52) in the main text for Project 1. \\

The code is sub-divided in the following way: First the objectives of the task are listed and a (i) bullet-point is exhibit when the item is fully solved. Starting by defining the initial parameters (numerical, uniform grid and data for hydrogen), the matrices of the linear system $M\phi=M\rho$ are build and later on the (Dirichlet) boundary conditions are imposed, increasing the degrees of freedom of the system by two (adding two additional row/columns). Once the density term defined with the appropriate change of basis $u(r),U(r),\phi(r)$ for clarity.\\

An example of running Sinajo with \emph{NN}=1000 (number of steps) and \emph{rmax}=6 (integration over six Bohr radius) is included below:\\ 

 {\fontfamily{qcr}\selectfont\
Sinajo (2022)\\
1.- Matricial system construction delay.\\
  Elapsed time is 0.021319 seconds.\\
2.- Does matrices M,L have the right \\number of non-zero values?\\
  Yes, only diagonal and adjacent \\  are non trivial.\\
3.- Solving the phi linear system\\ L phi = M rho.\\
  Elapsed time is 0.159059 seconds.\\
4.- Does the solution U(r) also fill the boundary conditions?\\
  Yes, U(0)=0.\\
  Yes, U(rmax)=1.\\
5.- The mean-squared error is \\  0.000000000.  Congrats!\\
}

 and in where the first and third sentences compute the elapsed time on the matrix definition and solution of linear system respectively. In the second bullet-point, a double-check of the non-trivial values of the sparse matrix is studied while in the fourth, the boundary conditions are re-check. Finally, the mean-squared error (MSE) is computed and if its less than the user tolerance \emph{tol} a congratulation message is displayed. \\

A figure of the previous run is displayed below.
\begin{figure}[!htbp]
    \includegraphics[width=8cm]{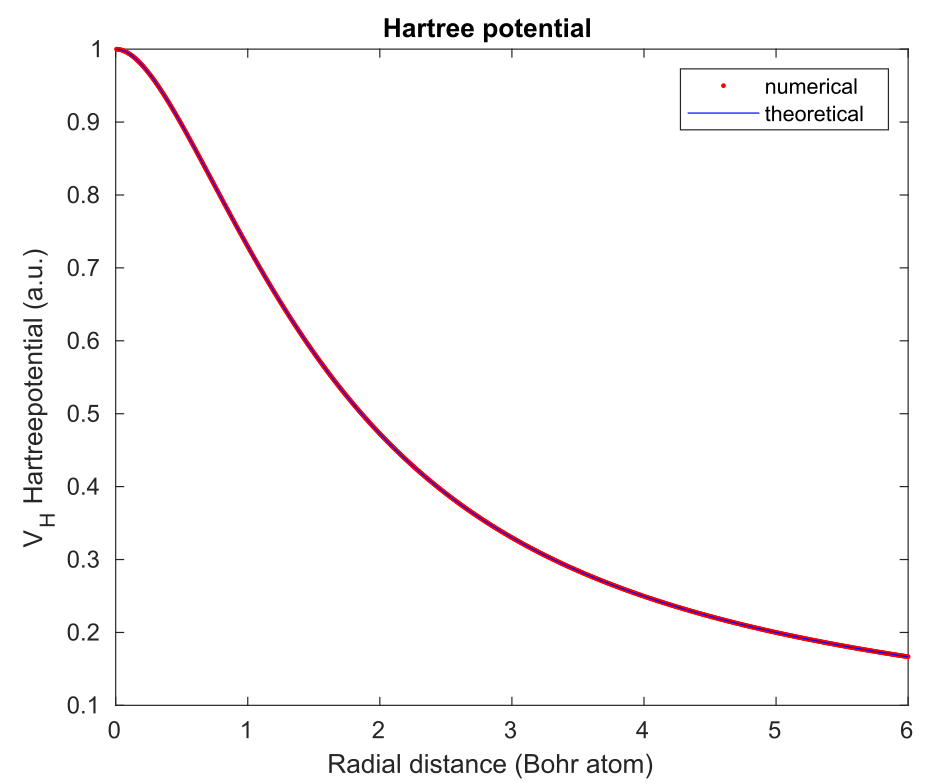}
    \caption{Poisson's equation in radial coordinates of the Hartree potential of the hydrogen $V_H=1/r-(1-1/r)e^{-2r}$.}
     \label{fig:concept}
\end{figure}

Finally, a study of the convergence is offered. We will study the two independent parameters: number of steps (NN), maximum radius value (rmax) and the corresponding dependent values: \ time elapsed and MSE of the solution,

\begin{center}
\begin{tabular} {||c|c||c|c|c||}
\hline
rmax & NN &$t_{matrix}(s)$ & $t_{linsys}(s)$ & MSE ($10^{-9}$)\\ 
\hline
\hline
\phantom{0}4 & 1000 & 0.021 & 0.1450 & \phantom{000}175\\
\hline
\phantom{0}4 & \phantom{0}100 & 0.001 & 0.0020 & \phantom{00}1380\\
\hline
\phantom{0}4 & \phantom{00}10 & 0.012 & 0.0080 & \phantom{0}82492\\
\hline
\phantom{0}6 & 1000 & 0.021 & 0.1590 & \phantom{00000}0\\
\hline
\phantom{0}6 & \phantom{0}100 & 0.009 & 0.0035 & \phantom{0000}51\\
\hline
\phantom{0}6 & \phantom{00}10 & 0.010 & 0.0016 & 275052\\
\hline
10 & 1000 & 0.026 & 0.1156 & \phantom{00000}0\\
\hline
10 & \phantom{0}100 & 0.010 & 0.0024 & \phantom{000}223\\
\hline
10 & \phantom{00}10 & 0.011 & 0.0024 & 117233\\
\hline
\end{tabular}
\end{center}

whose graphic representation leads to Figure 2.\\
\begin{figure}[!htbp]
    \includegraphics[width=8cm]{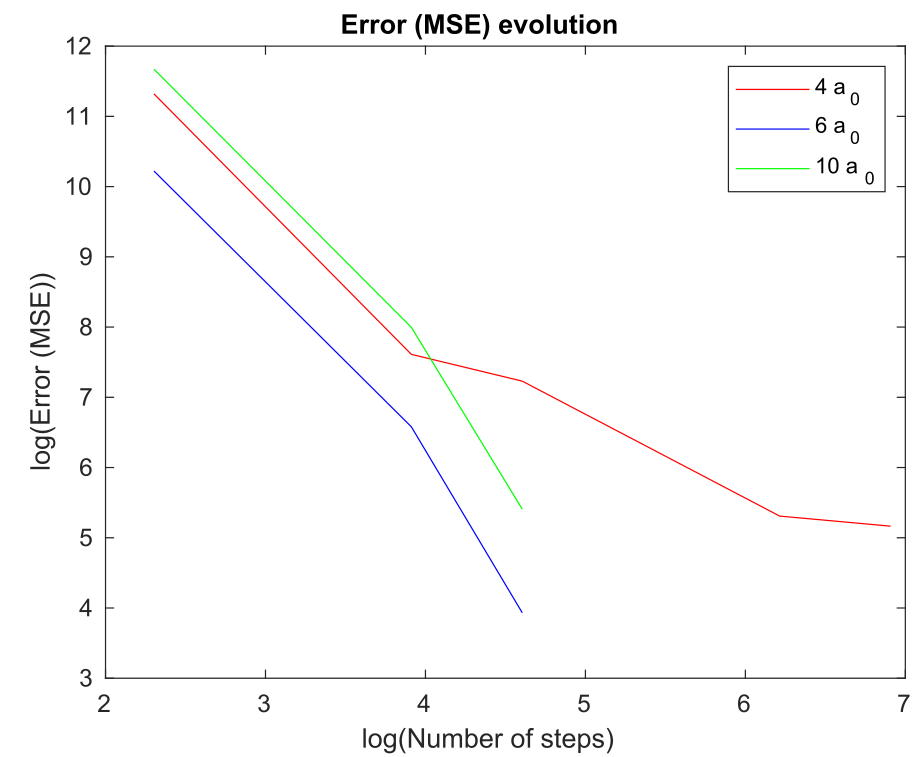}
    \caption{log-log relation between mean-squared error and number of steps for three different maximum radius, equivalent to $4,6$ and $10$ Bohr radius.}
     \label{fig:concept}
\end{figure}

From here, several results can be extracted and implemented for the next tasks: The maximum radius seems not to be a critical parameter so we can only consider a single case with $r_{max}\in(5,15)$. Also, the number of steps seems to be important but once few hundreds of points are studied, the solutions tends to converge. Finally, we can see that the time used in solving the linear system is the heavier, while sparse matrix filling exhibit a more chaotic (randomly distributed) with no linearity monotone progression.\\

As we have seen in Figures 2,3 the results meets the expectations as the blue theoretic line and the '+' sign of numerical simulation match. From here we can go towards the next section.

\section{Analytical contribution of nuclear potential in Kohn-Sham equation}
For this section, we just re-use the code implemented, and evaluating  using the basis $e[i]$ \\ 

 {\fontfamily{qcr}\selectfont\
e[i]:=Module\\
\{dxm =(x-xx[[i-1]])/(xx[[i]]-xx[[i-1]]), \\
dxp=1-(x-xx[[i]])/(xx[[i+1]]-xx[[i]])\}, \\
Piecewise[ \\ \{\{dxm,xx[i-1]] $\leq$ x $\leq$ xx[[i]]\}, \\
            \{ dxp,xx[[i]]<x<xx[[i+1]]\}\},0] \\
\\
}
we can evaluate the following integrals, once the grid is defined, with the Assumption: {$hh \geq 0$} s.t. $xx = {- hh, 0, hh, 2 hh, 3 hh, 4 hh, 5 hh, 6 hh, 7 hh...}$\\

 {\fontfamily{qcr}\selectfont\
Integrate[e[3]1/x e[3],\{x,0,Infinity\}\\
Integrate[e[4]1/x e[4],\{x,0,Infinity\}\\
Integrate[e[5]1/x e[5],\{x,0,Infinity\}\\
Integrate[e[6]1/x e[6],\{x,0,Infinity\}\\
Integrate[e[3]1/x e[4],\{x,0,Infinity\}\\
Integrate[e[4]1/x e[5],\{x,0,Infinity\}\\
Integrate[e[7]1/x e[6],\{x,0,Infinity\}\\
Integrate[e[5]1/x e[7],\{x,0,Infinity\}\\
}

leading to the following results:\\

\begin{figure}[!htbp]
    \includegraphics[width=5cm]{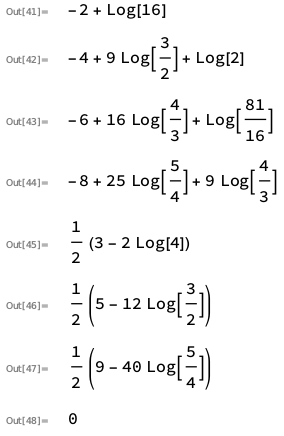}
     \label{fig:concept}
\end{figure}

\newpage

And whose general term can be directly encoded; except for the $(4,12,24...)$ which can be seen by successive differences or directly typing it into \emph{oeis.org} and getting the sequence \emph{A046092}: '4 times triangular numbers: $a(n) = 2*n*(n+1)$'.\\

Also, we can double-check expression (51) just using the following Assumption:\\$xx[[5]]>xx[[4]]>xx[[3]]>xx[[2]]>xx[[1]]>0$\\

 {\fontfamily{qcr}\selectfont\
Sum [1/xx[[i]]Integrate[e[i]×e[3]×e[3],\\
Sum [1/xx[[i]]Integrate[e[i]×e[4]×e[3],\\
Sum [1/xx[[i]]Integrate[e[i]×e[3]×e[5],\\
\{x,0,infinity\},\\
}
\begin{figure}[!htbp]
    \includegraphics[width=8cm]{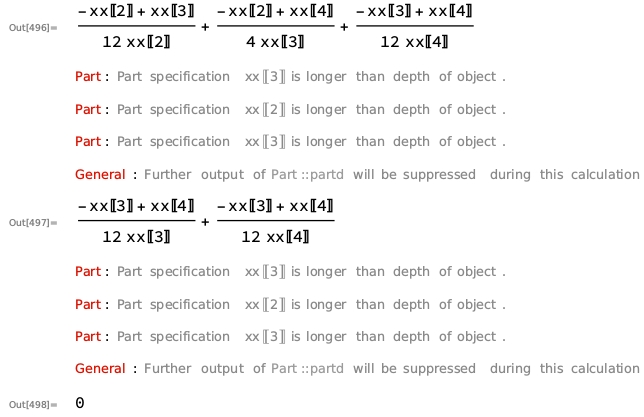}
     \label{fig:concept}
\end{figure}

confirming the values before mentioned. Both files are available in the supplementary material.

\section{Radial Schrodinger equation for case of the hydrogen atom}
With the previous ideas in mind we can try to solve the generalized eigenvalue problem for solving the radial Schrodinger equation (30) for the hydrogen, with MatLab \emph{eig}'s function and in which we can ignore the Hartree potential field and also the exchange-correlation terms as there is only one electron.\\

Using finite element method (FEM), we will compare the numerical result with the theoretical $E^{hydrogen}_0=-0.5$, ground state and $\psi_{hydrogen}=2e^{-r}/\sqrt{4\pi}$ wave function. \\

The code is again sub-divided in the following way: First the objectives of the task listed and defining the initial parameters similar to Task 1 (observe that the parameters for the hydrogen are mostly trivial), and where the matrices of the generalised eigenvalue system $(A/2+Q)\phi=\lambda S \phi$, similar to the usual problem but with the extra $S$ matrix and finally the eig-function problem is solved searching the minimun eigenvalue and obtaining its eigenfunctions and plotted.\\

An example of running Sinajo \emph{NN}=1000 (number of steps) and \emph{rmax}=8 (integration over eight Bohr radius) is included below:\\ 

 {\fontfamily{qcr}\selectfont\
Sinajo (2022)\\
1.- Matricial system construction delay.\\
  Elapsed time is 0.047676 seconds.\\
2.- Does matrices A,S have the right\\   number of non-zero values?\\
  Yes, only diagonal and adjacent\\ are non trivial.\\
3.- Solving the $\lambda$ generalized\\   eigenvalue problem.\\
  Elapsed time is 7.977270 seconds.\\
4.- The experimental ground state is \\      -0.517121700.\\
5.- The mean-squared error is\\       0.000049534.\\
}

 and in where, once again, the first and third sentences compute the elapsed time on the matrix definition and solution of generalised eigenvalue problem respectively, noting that this time the second part is quite more heavy to compute. In the second bullet-point, a double-check of the non-trivial values of the sparse matrix is studied while in the fourth, the ground state is shown. Finally, the mean-squared error (MSE) between the theoretical plane wave and numerical one (eigen-vector) is computed and if its less than the user tolerance \emph{tol} a congratulation message is displayed. This time, I couldn't obtain my 'congratulations' as when I tried with 2000 steps, the computer just display 'Out of memory'.\\

A graphic of the previous run is displayed in Figure 4, where both $u(r)$ (the eigen-function of (30)) and $\psi(r)$ are displayed. Finally, a study of the convergence is offered. We will study this time only the influence of the number of steps in the MSE of the solution, as in Section 1, we discuss that the other two parameters could be omitted.

\begin{figure} [!htbp]
    \includegraphics[width=7cm]{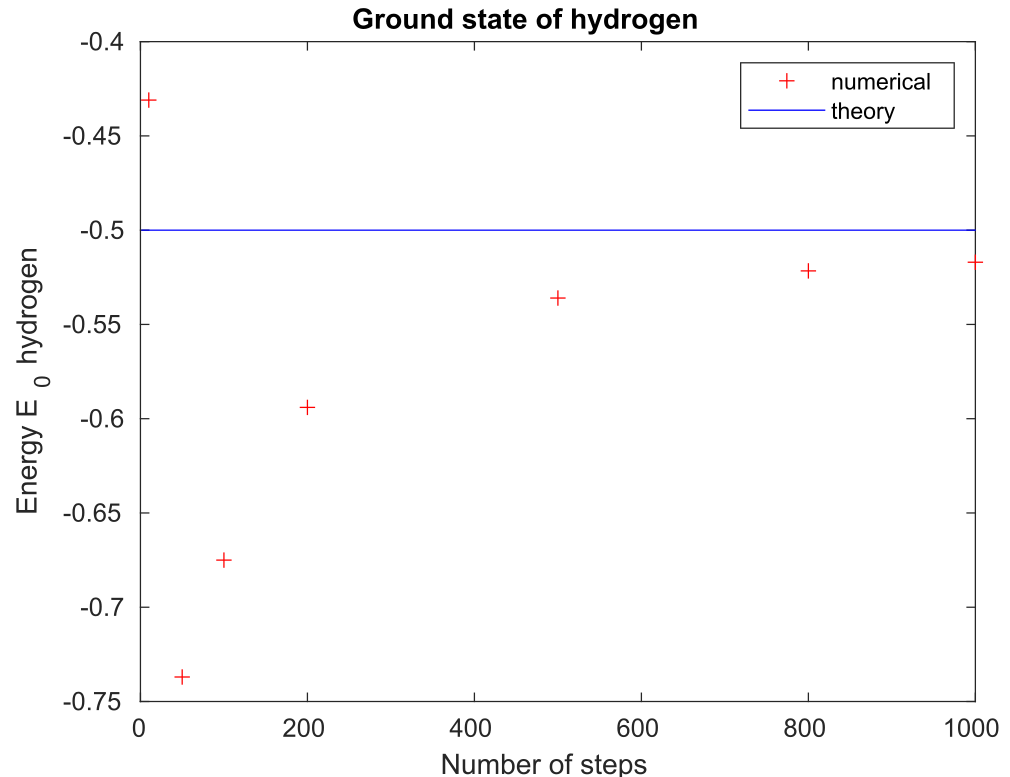}
    \caption{Converge of the numerical eigenvalue problem for ground state of hydrogen ($E_0=-0.5$ Hartree).}
     \label{fig:concept}
\end{figure}

It is interesting to see how with few points the energy is overestimated, but apart from that the converge is quite clear. We are ready to go straight-forward to the last section, where all the previous task will be use simultaneously.

\begin{figure*}[!htbp]
    \includegraphics[width=1\textwidth]{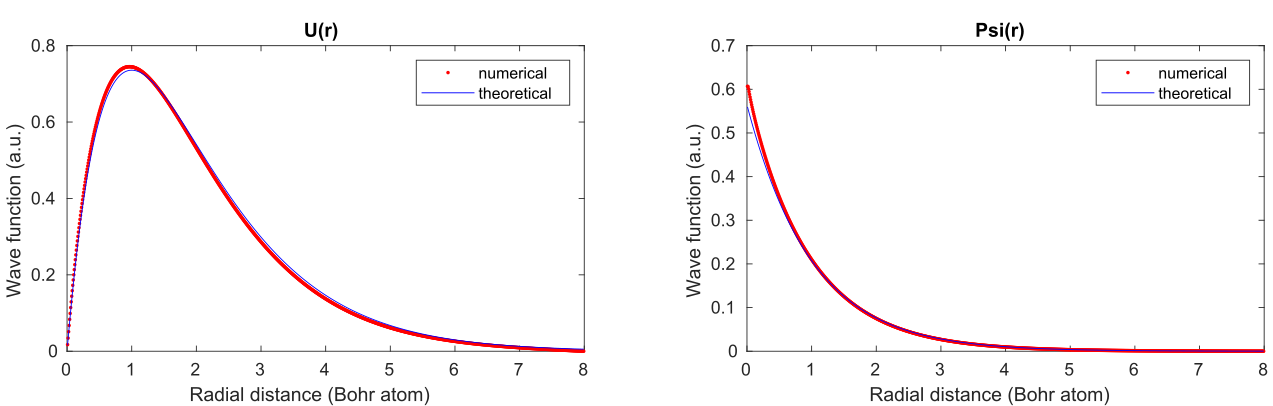}
    \caption{Kohn-Sham solutions or radial Schrodinger equations  of the hydrogen atom.}
\label{fig:poincare}
\end{figure*}

\section{Radial Schrodinger equation for the case of the helium atom}
If we already know how to compute the second derivative, the nuclear term and previously we can obtain the Hartree potential, with the help of equations (s4-s19) we can try to solve the generalized eigenvalue problem for solving the radial Schrodinger equation (30) for the helium, with MatLab builted function \emph{eig}'s and using (31) obtaining also the ground state value. Using, lastly, finite element method (FEM), we will compare the numerical result with the theoretical $E^{hel}_0=-2.903$ ground state and guessing the $\psi_{hel}$ wave function. \\

The code rely now on the sections: Final objectives, initial parameters for the helium and then we enter a loop, in which we are going to be as long as our solution is 'bad' (quantitatively, it means that our value of $E_0^{hel}$ differs more than $1\%$ of the theory and when we don't run the program more than \emph{iter-max} times). Once we are in, we need to obtain numerically all the potentials. Nuclear is already known from previous tasks. Finally for the exchange-correlation potential we need to perform some additional definitions.\\

From here, we again build the matrix system and solve the generalised eigenvalue problem. But now with our solution we are going to perform a self-consistency; we are going to run again the loop and change our initial parameters slightly (currently, $\alpha=0.1$) and executing automatically the program again.\\

An example of running Sinajo with \emph{NN}=1000 (number of steps) and \emph{rmax}=6 (integration over ten Bohr radius) is included below:\\ 

 {\fontfamily{qcr}\selectfont\
Sinajo (2022)\\
1.- Self consistency iteration.\\
2.- The experimental ground state is CURRENTLY -2.797568686 obtained in 1 steps.\\
2.- The experimental ground state is CURRENTLY -2.802715288 obtained in 2 steps.\\
...\\
2.- The experimental ground state is CURRENTLY -2.822842453 obtained in 24 steps.\\
3.- The experimental ground state is   FINALLY -2.822820430 obtained in 25 steps.\\
         Congrats!\\
}

at this time, we only focus on the time elapsed between different trials to see if there exists some tendency and in part 2 we display the approximate solution obtained (that, in principle, has an error less than 3\%) and the number of iterations needed.\\

A figure of the previous run is displayed below, where both the density function (from (30)) and wavefunction are displayed.
\begin{figure*}[!htbp]
    \includegraphics[width=0.8\textwidth]{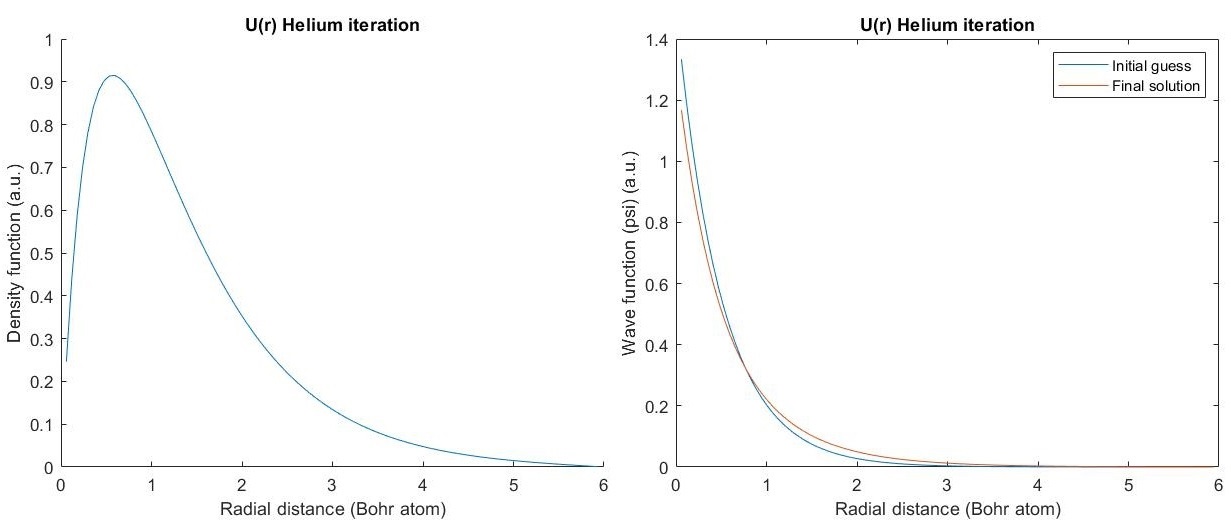}
    \caption{Kohn-Sham solutions or radial Schrodinger equations  of the helium atom.}
\label{fig:poincare}
\end{figure*}

To finish, a study of the convergence is offered below. There are many parameters to study; in particular it is quite interesting to see the changes that happens when we change the initial 'seed' (the initial wave-function guess) and the numerical parameter $\alpha$. In our case, the first parameter was studied introducing some perturbations to the initial state (we add some noise and also multiply the wave-function itself for a certain constant). The results weren't definitely as in many cases the convergence was or broken or too slow (we always consume the maximum iterations). Nevertheless we estimate that if the perturbation don't move the initial state more than $5\%$ (that means that the global wave-function is multiplied or divided by $1.05$ or the maximum Gaussian noise is $1/20$ of the minimum value), it still converges. With respect to the parameter $\alpha$, we have the same behavior, trials with $(0.5,0.01, 0.25)$ were performed with distinct rate of converging, which leads to different number of steps for obtaining an accurate result (with an error less than the tolerance imposed). Nevertheless, the difference with the theoretical value is based on the different approaches (Density Functional Theory with Kohn-Sham eqn. and Local Density Approximation.) \\

\section*{Discussion}
We have approximate to an accurate value the ground states and wave-functions for \emph{hydrogen} and \emph{helium} atoms. Although exchange-correlation terms seems to be tough, it would be very interesting to study the average weight of each of the potential for every region, in particular to see the influence of them, as representing the electronic cloud or the nuclei-electron interaction. Also, it is possible to simulate other simple atoms like \emph{lithium} to see the advantages and weakness of the numerical DFT-LDA method. An additional file corresponds to the convergence plots.\\

This code could be useful in undergraduate studies for a deeper understanding of numerical simulations and approximation series, while demonstrating the powerful tool that we have in our hands to simulate accurate historically-challenging physical problems. In further reviews of the software, we are planning to develop as an open-source toolbox for Condense Matter simulations $\blacksquare$.

\section*{Bibliography}
\section*{Supplementary material}

\end{document}